\documentclass[preprint2]{aastex}

\usepackage{natbib}
\usepackage{graphicx}
\usepackage{multirow}
\usepackage{txfonts}

\begin{document}

\shorttitle{Properties of MAXI~J1543-564 with TCAF and POS Models}
\shortauthors{Chatterjee et al.}

\title{Accretion Flow Properties of MAXI~J1543-564 During 2011 Outburst from TCAF Solution}
\author{Debjit Chatterjee\altaffilmark{1}, Dipak Debnath\altaffilmark{1,2}, Sandip K. Chakrabarti\altaffilmark{3,1}, Santanu Mondal\altaffilmark{1,4}, Arghajit Jana\altaffilmark{1}}
\altaffiltext{1}{Indian Center for Space Physics, 43 Chalantika, Garia St. Rd., Kolkata, 700084, India.}
\altaffiltext{2}{Institute of Astronomy, National Tsing Hua University, Hsinchu, 30013, Taiwan.}
\altaffiltext{3}{S. N. Bose National Centre for Basic Sciences, Salt Lake, Kolkata, 700098, India.}
\altaffiltext{4}{Instituto de F\'isica y Astronom\'ia, Facultad de Ciencias, Universidad de Valpara\'iso, Gran Bretana N 1111, Playa Ancha, Valparaíso, Chile}

\email{debjit@csp.res.in; dipak@csp.res.in; chakraba@bose.res.in; santanu@csp.res.in, argha@csp.res.in}

\date{Received 2016 March 14; accepted 2016 May 11 (in ApJ Main)}

\begin{abstract}
We derive accretion flow properties of the transient black hole candidate (BHC) MAXI~J1543-564 
using the RXTE data. We use Two-Component Advective Flow (TCAF) solution to fit the data of the 
very initial rising phase of outburst (from 2011 May 10 to 2011 May 15). $2.5-25$~keV spectra 
are fitted using the TCAF solution {\it fits} file as a local additive table model in XSPEC. We 
extract physical flow parameters such as the two component (Keplerian disk and sub-Keplerian halo) 
accretion rates and size and the property of the Compton cloud (post shock region close to 
a black hole). Similar to other classical transient BHCs, monotonic evolution of low frequency 
quasi-periodic oscillations (QPOs) are observed during the rising phase of the outburst, which 
is fitted with the propagating oscillatory shock (POS) model which describes how the Compton cloud 
properties change from day to day. From the nature of variations of TCAF model fitted physical 
flow parameters and QPOs, we only found hard-intermediate and soft-intermediate spectral states 
during this phase of the outburst under study. We also calculate frequency of the 
dominating QPOs from the TCAF model fitted shock parameters, and found that they roughly match with 
the observed and POS model fitted values. From our spectro-temporal study of the source with TCAF 
and POS models, the most probable mass of the BHC is found to be $12.6-14.0$~$M_\odot$, 
or $13^{+1.0}_{-0.4}~M_\odot$.

\end{abstract}


\keywords{X-Rays:binaries -- stars individual: (MAXI J1543-564) -- stars:black holes -- accretion, accretion disks -- shock waves -- radiation:dynamics}

\section{Introduction}

Compact objects such as black holes (BHs) do not emit radiation by themselves. They can be 
detected by electromagnetic radiation emitted by accreted matter falling on them. Most of the 
black hole candidates (BHCs) are observed in close binaries in our Galaxy. 
They accrete matter from their companions through Roche lobe flow and from the winds. 
Some black hole candidates are transients in nature. Outbursts of these 
transient X-ray binaries exhibit daily variation of temporal and spectral 
properties and each of these observations gives us an opportunity to understand the 
accretion processes around the respective black hole from detailed 
spectral and temporal analysis. For this, one requires to have a realistic solution of the flow 
and its radiative properties which preferably has minimum number of parameters. 
Sizable number of scientific 
papers are available in the literature from many groups (e.g., McClintock \& Remillard, 2006; 
Debnath et al., 2008, 2013; Tomsick et al., 2014) to model observations. In general, these candidates 
show hard (HS), hard-intermediate (HIMS), soft-intermediate (SIMS) and soft (SS) spectral states during 
any particular epoch of an outburst (see, Debnath et al. 2013 and references therein). High and low frequency 
quasi-periodic oscillations (QPOs) have also been observed in their power density spectra (PDS) in some of 
these spectral states (see, Remillard \& McClintock, 2006 for a review).

MAXI~J1543-564 was first discovered by MAXI/GSC on 2011, May 8 (Negoro et al. 2011) at  R.A.=$15^h43^m9.12^s$, 
Dec.=$56^\circ25'15.6''$. The outburst of this source has been extensively observed by MAXI and Swift 
(e.g. Kennea et al. 2011), RXTE (e.g. Altamirano et al. 2011). Munoz-Darias et al. (2011) confirms the source 
as a potential BHC from their spectral and timing analysis. The nature of the companion (or, companions) is not 
confirmed since there was no significant variability in optical emission (Russell et al. 2011; Rau et al. 2011; 
Rojas et al. 2011). Stiele et al. (2011) estimated a minimum distance of the source to be $8.5$~kpc.
Miller-Jones et al. (2011) reports a weak radio emission on MJD=55695.73 (2011 May 14), which is consistent 
with the prediction made by Kennea et al. (2011) and Munoz-Darias et al. (2011) that the candidate made a 
transition from HIMS to SS between 2011 May 13 (MJD=55694.09) and 2011 May 15 (MJD=55696.65).
 
To study the flow properties of an outbursting BHC, one requires a solution which provides the mass, 
accretion rates and size of the Compton cloud from the observed photon spectrum on each day. Recently, 
after inclusion of the Two-Component Advective Flow (TCAF) model (Chakrabarti \& Titarchuk, 1995, hereafter CT95), 
i.e., producing a model fits file using a very large number of theoretical spectra, into HEASARC's spectral 
analysis software package XSPEC (Arnaud, 1996) as a local additive table model, we found that TCAF is capable 
of extracting physical parameters of the flows on a daily basis (see, Debnath et al., 2014, 2015a,b; 
Mondal et al., 2014; Jana et al., 2016, hereafter DCM14, DMC15, DMCM15, MDC14, JDCMM16 respectively). 
TCAF model fits extract two component (Keplerian disk and sub-Keplerian halo) accretion rates, 
shock location (outer edge of the Compton cloud) 
and compression ratio from each observation. Even one can obtain independent estimation of the probable 
mass from each observations. Combined together from the observational set, a reasonable mass of the
BH from TCAF model fits (Molla et al., 2016) can be obtained. One can also have an idea of the observed 
frequency of the dominating QPOs (if present; see DCM14), viscous time scale (see, JDCMM16), etc. from the 
TCAF model fitted/derived physical flow parameters since it is considered to be a resonance oscillation of 
the Compton cloud boundary (i.e., the shock). Properties of different spectral states, and their transitions 
could also be explained from the nature of the variations of accretion rate ratio (ARR; ratio between halo 
to disk rates) and QPOs (if present). 

Low frequency QPOs are commonly observed in hard and intermediate spectral states of transient BHCs. 
Generally, it has been observed that frequency of these QPOs monotonically increases with time (day) 
during rising HS and HIMS, and decreases with time during HIMS and HS of the declining phases of 
an outburst of a transient BHC. The evolutions of the observed QPO frequencies are explained with the 
propagating oscillatory shock (POS) model (Chakrabarti et al., 2005, 2008; Debnath et al., 2010, 2013; 
Nandi et al., 2012). According to POS, QPOs occur due to resonance between cooling and infall time of 
the post-shock region (Molteni et al., 1996; Chakrabarti et al., 2015) or due to non-satisfaction of 
the Rankine-Hugoniot conditions (Ryu et al., 1997). According to POS, frequency of the QPOs is inversely 
proportional to the infall time of the advective flow in the Compton cloud. From the model fitted QPO evolution, one 
can get the shock location, velocity, compression ratio, etc. These shock parameters, also could be 
verified with TCAF model fitted spectral parameters, since we are using same shock in both the cases. 
Thus connectivity of the day to day variation of TCAF comes from POS model.

The successful interpretation of accretion flow dynamics and QPO evolutions with TCAF and POS models respectively, 
motivated us to study early rising phase of the 2011 outburst of MAXI~J1543-564 with these two models. 
In the next Section, we briefly describe data analysis technique using HeaSoft package. In \S 3, we present spectral 
and temporal analysis results of the source with both TCAF and POS models. Here, we also calculate QPO frequencies 
obtained from the TCAF model fitted shock parameters (location and compression ratios) and compare them with observed 
and POS model fitted. We also estimate a most probable range of the mass for this BHC from two methods discussed 
in Molla et al. (2016). Finally in \S 4, we conclude our understanding of the accretion flow properties of 
this BHC during its very early phase of 2011 outburst from the TCAF model fit.

\section{Observation and Data analysis}

RXTE observed the source immediately after two days of its discovery roughly on a daily basis starting from 
2011 May 10 to 2011 September 30. We analyze first seven observations of the RXTE Proportional Counter Array (PCA) 
instrument in the rising phase of the outburst, starting from 2011 May 10 (MJD=55691.09) to 2011 May 15 (MJD=55696.66). 
For spectral and timing analysis, we follow the standard data extraction and analysis methods as defined in 
Debnath et al. (2013, 2015a,b) using HEASARC's software package HeaSoft version HEADAS 6.16 and XSPEC version 12.8. 

PCA spectra in the $2.5-25$ keV energy band are fitted with the current version (v0.3) of the TCAF model {\it fits} 
file, as an additive table model in XSPEC, which needs to supply five model input parameters, namely, 
$i)$ black hole mass ($M_{BH}$) in solar mass ($M_\odot$) unit, $ii)$ Keplerian accretion rate ($\dot{m_d}$ in 
Eddington rate $\dot{M}_{Edd}$), $ii)$ sub-Keplerian accretion rate ($\dot{m_h}$ in $\dot{M}_{Edd}$),
$iv)$ location of the shock ($X_s$ in Schwarzschild radius $r_g$=$2GM_{BH}/c^2$), and $v)$ compression ratio 
(R=$\rho_+ / \rho_-$, where $\rho_+$ and $\rho_-$ are the post- and pre-shock densities respectively) of the shock. 
For the strongest shock, $R$ could be $4-7$, depending on the polytropic index of the flow. 
In our case of a hot and rotating advective flow, the flow is not highly supersonic and thus we have weaker shocks. 
The model normalization ($N$) is not required explicitly, since it comes out through the fit, and all we require is that 
it remains in a narrow range during the entire phase of our observations. It is difficult to put $N$ as a single factor, 
since the Compton cloud is not in the same plane as the Keplerian disk (unlike in disk black body model, 
where entire flow components are assumed in the same plane and a single inclination angle appears in normalizing 
the whole spectrum). In any case, $N$ is a function of constant (albeit unknown) physical parameters, 
such as mass of the BH, distance `D' and the inclination angle `i'. 
For all observations, we assume the constant Hydrogen column density ($N_H$) 
at $0.9 \times 10^{22}~atoms~cm^{-2}$ (Kennea et al., 2011) for 
photon absorption model {\it phabs} and a fixed 1\% systematic instrumental error. To get a better fit, 
we use an additional Gaussian line of energy $\sim 6.5$~keV for Iron emission line. After achieving the 
best fit based on reduced chi-square value ($\chi^2_{red} \sim 1$), the XSPEC command `err' is used to 
find 90\% confidence positive and negative error values for the model fitted parameters. In Table I, average 
values of these two $\pm$ errors are mentioned in the superscripts of the model fitted parameter values. 

We looked for low frequency QPOs, after generating PDS using ``powspec" task of the XRONOS package. This task 
computes rms fractional variability on $2-15$~keV Proportional Counter Unit 2 (PCU2; including all six layers) 
light curves of $0.01$ sec time bin. These light curves are generated using the PCA {\it Event} mode data of 
a maximum timing resolution of $125\mu s$. To find centroid frequencies of the QPOs, PDS are fitted with 
Lorentzian profiles and ``fit err" command is used to get $\pm$ error limits. The monotonic evolution of the 
QPOs during the initial five observations of the rising phase of the outburst is fitted with the POS model. 
In Table 2, we present the POS model fitted parameters (instantaneous shock location, velocity, compression ratio) 
along with QPO frequencies calculated/predicted using POS and TCAF models.

\section{Results}

We make a detailed spectral and temporal study of the initial seven RXTE/PCA observations from the 
rising phase of the very first (2011) outburst of MAXI~J1543-564 after its discovery on 2011 May 8. 
TCAF model fitted/derived spectral parameters are given in Table 1 and POS model fitted shock 
parameters along with observed or predicted QPOs (with POS and TCAF models) are given in Table 2.

In Figure 1, we show 2011 MAXI~J1543-564 outburst profile as observed by MAXI satellite in the energy 
range of $2-10$~keV. The region of the RXTE/PCA observations (from the rising phase of the outburst), which 
are presented in the current paper, is marked by the region between two arrows. In Figure 2(a-b), we show TCAF 
model fitted two spectra, selected from two different observed spectral states, HIMS (Ris.) and SIMS (Ris.) respectively. 
In Figure 3, variations of TCAF model fitted parameters, PCU2 rate and observed QPOs during the
initial rising phase of the outburst is shown. In Figure 3(a-d), variations of Keplerian 
disk rate $\dot{m_d}$, sub-Keplerian halo rate $\dot{m_h}$, shock location $X_s$, and 
compression ratio $R$ are shown. The variations of the total flow rate ($\dot{m_d}$+$\dot{m_h}$), 
ARRs are shown in Figs. 3(f) and (g) respectively. In Figs. 3(e) and (h), variations of $2-15$~keV 
PCU2 count rate and observed dominating QPO frequencies are shown respectively. 
In Fig. 4(a), we show POS model fitted monotonic evolution (increasing) of the QPO frequencies 
(type-C) for the five observations of MAXI~J1543-564 which belong to the hard intermediate state. 
In Figure 4(b), variations of the shock locations and compression ratios, obtained from POS model 
fit are also shown. In Fig. 5(a-b), we show the variations of the TCAF model normalization and 
mass of the BH. In Fig. 5(c), variation of the POS model fitted $\chi^2_{red}$ with BH mass is shown. 

\subsection{Spectral Evolution of MAXI J1543-564: Analysis with TCAF Solution}

Evolution of spectral properties and accretion flow dynamics around the BH during initial rising phase 
of the very first outburst of MAXI~J1543-564 are clear from our analysis. From the variation of the fitted
flow parameters and nature of QPOs (if present), we discover only two spectral states, such as, HIMS and SIMS. 
We also note that on the first RXTE/PCA observed day (2011, May 10), the source is already in HIMS, although 
it was discovered on 2011, May 8. Since outbursting is unpredictable, it is not unusual to miss the initial 
hard states (see also, DMCM15 for 2010 outburst of MAXI~J1659-152).

\noindent{\bf HIMS (Rising):}
From the variations of PCU2 count and total flow rates, it is clear that as the day progresses, more 
matter from the companion reaches the black hole which increased the total X-ray intensity. 
Accretion flow dynamics becomes clear when we look into the variation of the rates of the two components, 
namely, the Keplerian disk ($\dot{m_d}$) and the sub-Keplerian halo ($\dot{m_h}$), outer boundary of the Compton
cloud (i.e., shock location $X_s$) and the compression ratio ($R$) parameters. On the first day of our observation, 
there is a clear dominance of halo rate ($\dot{m_h}=0.304$$\dot{M}_{Edd}$) over disk rate 
($\dot{m_d}=0.022$$\dot{M}_{Edd}$) which maintains a moderately strong ($R=2.76$) shock far away ($X_s=215~r_g$) 
from the BH. As the day progresses, the shock becomes weaker and moves towards the BH horizon with 
a rise in the disk rate and decrease in the halo rate. As a result of that, a decrease in accretion rate ratio (ARR)
and monotonic increase in dominating QPO frequency are observed. 
On the fifth observation day (2011 May 13; MJD=55694.89), $\dot{m_h}$ reaches its minimum value 
($\dot{m_h}=0.156$$\dot{M}_{Edd}$) with maximum observable (monotonically evolving) QPO frequency 
($5.70$~Hz), low ARR value ($=0.32$). On this particular day, outer boundary of the
Compton cloud (i.e. $X_s$) reaches $\sim 92~r_g$ but the shock is weakened due to cooling by high Keplerian rate. 
After this date only a weak centrifugal barrier can form, but no sharp shock boundary can form. 
This observation thus signifies the transition between two spectral states.

\noindent{\bf SIMS (Rising): }
The last two observations of our analysis fall in this spectral state. Variations of QPOs (sporadic) 
and TCAF model fitted parameters are similar to those seen in other classical objects 
(see, MDC14, DMC15, DMCM15). We observe a $5.08$~Hz QPO on the first 
day and no QPO on the last day. During this phase of the outburst, 
shock becomes much weaker ($R \simeq 1.05$) and is located in the same effective distance ($X_s$). 
A clear dominance of the Keplerian disk rate over the halo rate is observed, which results in a low ARR. 

\subsection{Evolution of QPOs with POS model}

In general, low frequency QPOs are observed during hard and hard intermediate spectral states of BHCs. This is because 
quasi-periodic variation post-shock region due to shock oscillation and the resulting oscillation of Comptonized X-ray 
intensity (Chakrabarti \& Manickam, 2000) . It has been observed in several transient BHCs that type-C QPOs (generally 
observed in HS and HIMS) evolve with time (day) during rising and declining phase of the outbursts and type-A or B QPOs 
are observed sporadically on and off during SIMS (see, Debnath et al., 2008, 2013; Nandi et al., 2012). 
Although not everyone agree on the origin of these QPOs, but according to TCAF, these are easily explained by 
the resonance oscillation (type-C), weak oscillation (type-B) of the  Compton cloud boundary (i.e. shock) 
or even of the shock-free centrifugal barrier (type-A).
For type-C QPOs, shock oscillation may occur due to fulfillment of the resonance condition between cooling 
and infall time of the post-shock region (Molteni et al., 1996; Chakrabarti et al., 2015) or due to the non-fulfillment
of the Rankine-Hugoniot conditions (Ryu et al., 1997). The frequency of the QPOs is inversely proportional to the 
infall time from the location of the shock (i.e., outer boundary of the `CENtrifugal pressure dominated BOundary Layer', 
or CENBOL, which acts as the `Compton cloud', Chakrabarti \& Manickam, 2000).

In the initial six PCA observations (2011 May 10 to 14, i.e., MJD=55691.09 to 55695.67), similar to other 
classical transient BHCs, evolution of type-C QPOs (from $1.05 - 5.70$~Hz) is observed in the first five days 
of the outburst. On the sixth day, observed QPO ($5.08$~Hz) is of type-B. We study evolution of the QPO frequency 
with the same POS model as in the earlier objects (Chakrabarti et al., 2005, 2008; Debnath et al., 2010, 2013; 
Nandi et al., 2012). According to POS, during the rising phase of the outburst, the shock moves in while monotonically 
reducing shock strength (as cooling increases due to increase in Keplerian rate) and it results in monotonic
rise in the QPO frequency. In contrast, during the declining phase of the outburst, the shock moves away 
from the BH since matter supply from the companion reduces resulting in monotonic decrease in QPO frequency. 
For the sake of completeness, we again discuss governing equations of the POS model. 

The equation to find QPO frequency is given by (Chakrabarti \& Manickam, 2000; Chakrabarti et al., 2005), 
$$
\nu_{QPO} =\beta/[X_s (X_s-1)^{1/2}], 
\eqno{(1)}
$$
where, the shock strength 
$$\beta =1/R_0 \pm t_{d}^2\alpha,\eqno{(2)}$$
`$R_0$' is the value of compression ratio $R$ on the first day (0$^{th}$ day), $t_{d}$ is the time in days, 
and $\alpha$ is a constant number, which decides how $R$ becomes stronger or weaker with the QPO evolution period. 
The instantaneous shock location is defined as, 
$$
X_s(t)=X_{s0} \pm V t/M_{BH} ,
\eqno{(3)}
$$
where, $X_{s0}$ is the shock location on the first observation. The shock could be accelerating or decelerating. 
The instantaneous shock velocity is defined as 
$$ V(t)= V_{0} \pm f t_{d}, \eqno{(4)}$$
where, $V_0$ is the velocity of the first observation and `$f$' is the acceleration/deceleration. Here '+ve' 
sign is for accelerating shock and '-ve' sign is for decelerating shock wave propagation.

POS model fitted parameters of the QPO evolution during initial five days (2011, May 10-13, i.e., 
MJD=55691.09 to 55694.89) of our observation are presented in Table 2. We get best fitted 
($\chi^2_{red}=0.90$) evolution for the mass of the BH $M_{BH}$ as $13~M_\odot$. According to POS, 
the shock starts to move towards BH from $\sim 210~r_g$ with an initial velocity of $\sim 2450~cm/sec$ 
and a deceleration ($f=-345~cm/sec/day$). It reaches $\sim 114~r_g$ on the last day of the QPO evolution. 
During this phase of the outburst, POS model calculated $R$ reduces from $\sim 2.44$ to $\sim 1.08$ 
with constant $\alpha=0.037$, which roughly matches with TCAF model fitted $R$ values (see, Col. 4 of Table 1). 

\subsection{Prediction of BH Mass with TCAF and POS model fits}

Molla et al. (2016) estimated mass of the BHC MAXI~J1659-152 with the spectro-temporal analysis methods, 
which motivated us to estimate the probable mass range of MAXI~J1543-564. They used two methods, one is 
the TCAF model fitted {\it $i)$ constant normalization parameter} method, and the other is 
{\it $ii)$ by studying evolution of the QPOs with the POS model}. TCAF model normalization (N) 
being a factor which only depends on the mass and distance of the black hole and the inclination angle $`i'$ 
of the orbital plane, should not vary on a daily basis, unless the disk precesses and the projected surface 
has a variable emission area or there are significant outflow activities which is not included in 
TCAF fits file.

While fitting $2.5-25$~keV PCA spectra with TCAF, we kept all the parameters free, and found that model 
normalization and mass come in narrow ranges of $\sim 11.5-12.8$, and $\sim 13.5-14.0$~$M_\odot$ respectively 
(see, Col. 7 \& 6 of Table 1). The variations of these two parameters are shown in Fig. 5(a-b). 
Similarly, when we are fitting evolution of monotonically increasing QPOs during initial five days 
of our analysis, we obtained the best fit using $M_{BH} = 13~M_\odot$. Now, we varied mass of the BH in the 
POS equation to see the deviations of the model fitted $\chi^2_{red}$ from the best fitted 
value (=0.90). This variation of $\chi^2_{red}$ with mass is shown in the Fig. 5(c). If we restrict 
ourself to the $\chi^2_{red}$ value of $\leq 2.7$ (90\% confidence) for the best fits, we get probable 
mass range of the source as $\sim 12.6-13.6$~$M_\odot$. Combining these two methods, we determine the mass of 
the BHC to be in the range of $\sim 12.6-14.0$~$M_\odot$ or $13^{+1.0}_{-0.4}$~$M_\odot$. 
Our preferred mass is $13$~$M_\odot$ since it is the POS model fitted mass value.

\subsection{Prediction of QPOs with TCAF Model}

It is well known that the spectral and timing properties in BHCs are strongly correlated to each other 
as the location of the shock controls both the properties (CT95; Chakrabarti \& Manickam, 2000; Debnath et al. 2013). 
This correlation is thus intrinsic to the TCAF solution 
since the spectral and timing features are the outcome of the solution of the 
same set of transonic flow equations. In JDCMM16, a hysteresis diagram namely 
{\it accretion rate ratio intensity diagram (ARRID)} is plotted between ARR vs. 
PCA count rate, where different spectral states are found to be 
correlated with different branches of the diagram. In DCM14, dominant frequencies of the QPOs for three 
different BHCs (H~1743-322, GX~339-4, and GRO~J1655-40) were predicted using TCAF model fitted shock parameters. 
This could be done because the same shock parameters (namely, $X_s$, and $R$) which are used to define 
the size of the Compton cloud and matter densities in pre- and post-shock regions, is also used to find 
QPO properties in POS model. When the shock (outer boundary of CENBOL) oscillates, size of the 
Compton cloud varies, causing a variation of X-ray intensities in the light curve. As a result of that 
we observed QPOs in PDS. According to TCAF, using Eq. (1), the frequency of the QPOs could be calculated 
with model fitted shock parameters ($X_s$ and $R$), if mass of the BH is known.


We calculate frequency of the dominant type-C QPOs observed in the first five days of the outburst keeping 
mass of the BH as $13~M_\odot$ and found that for the initial four days, it roughly matches with the observed 
and POS model fitted values and on the fifth day a large deviation of $\sim 2.0$~Hz from the observed one 
is observed (see, Col. 9 of Table 2), indicating possible deviation from resonance condition which was used
in determining the shock oscillation frequency.

\section{Discussion and Concluding Remarks}

We study evolution of both spectral and timing properties during very early phase of the very first outburst 
of MAXI~J1543-564 after its discovery on 2011 May 8 (MJD=55689). To infer accretion for dynamics of the source, 
we use RXTE/PCA data from 2011 May 10 to 15. Spectra are fitted with the current version (v0.3) of the TCAF 
model {\it fits} file (see, DMC15) to extract physical flow parameters (e.g., Keplerian and sub-Keplerian 
flow rates, shock locations, and compression ratios, mass, etc.) directly from spectral fits (see, Table 1). 

From the nature of the variations of ARRs and QPOs, we observe only two spectral states, HIMS and SIMS, 
which is consistent with the results of Kennea et al. (2011) and Munoz-Darias et al. (2011). We observe 
a transition between these two spectral states on the fifth observation day i.e., 2011 May 13 (MJD=55694.89), 
since on this particular observation, the maximum frequency of the monotonically increasing QPO is observed. 
Also, on this day, the ARR reaches at a very low value. On the first day, we observe a high halo rate as 
compared to the disk rate and as day progresses, the shock moves in due to the shrinking of the Compton cloud 
i.e., CENBOL. The CENBOL size is reduced since during these days, ARR decreases from $13.8$ to $0.32$ due to 
the increase in disk rates and decrease in halo rates. Last two observations of our analysis belong to SIMS, 
since the QPO frequency decreases on the sixth observation and no QPO was present on the seventh observation. 
This sporadic nature is the signature of SIMS (see for e.g., Nandi et al., 2012; Debnath et al., 2013). 
A roughly constant low ARR values is also observed in these two observations, which is also consistent with 
the TCAF model fitted results for other objects (MDC14, DMC15, DMCM15).

Type-C QPOs during initial five observations of our analysis are independently fitted with POS model to 
find instantaneous shock location, compression ratio, velocity of the propagation, etc. (see, Table 2). 
This is the same model which was used to study monotonic evolutions of QPO frequencies during rising and 
declining phases of the outbursts of few other classical transient BHCs by our group (for e.g., GRO~J1655-40, 
XTE~J1550-564, GX~339-4, H~1743-322, MAXI~J1659-152, IGR~J17091-3624). We compare POS model fitted shock 
parameters ($X_s$ and $R$) with that of the TCAF model fitted spectral results and found them to be roughly 
consistent. According to POS, during the evolution period of $\sim 4$~days, shock location ($X_s$) changed 
from $\sim 210~r_g$ to $\sim 114~r_g$ and the shock strength is progressively weakened. This is due to loss 
of heat and pressure from the post-shock region due to inverse Comptonization as Keplerian rate is increased.

We continued our analysis by estimating the probable mass of the black hole by using methods given in
Molla et al. (2016). In the constant TCAF model normalization method, we kept all the 
TCAF parameters as free while fitting spectra and 
found narrow variations of the model normalization ($\sim 11.5-12.8$), and mass ($\sim 13.5-14.0$~$M_\odot$). 
Similarly, to fit QPO frequency evolution with POS model, we supplied the mass of the BH and found the best fit 
for $M_{BH}$ to be $13~M_\odot$. To get the best fitted ($\chi^2_{red} \leq 2.7$) mass range, we refitted QPO evolution 
by varying $M_{BH}$ and found the probable mass range to be $\sim 12.6-13.6$~$M_\odot$. Combining these two methods we get 
the mass of MAXI~J1543-564 in the range of $\sim 12.6-14.0$~$M_\odot$. 

We also calculated frequency of the dominant type-C QPOs using TCAF model fitted shock parameters 
($X_s$ and $R$) as in DCM15. According to Molteni et al. (1996) and Chakrabarti et al. (2015), 
the shock location coming from TCAF fits, could be used to calculate the frequency of the QPOs 
after applying resonance condition. For the initial four days, calculated QPO frequency roughly 
matches with the observed values, and POS model fitted values, but on the fifth day we observe 
a large deviation, possibly because of the deviation from resonance condition when HIMS ends.

It is to be noted that our TCAF code uses Paczy\'nski-Wiita potential (Paczy\'nski \& Wiita, 1980)
as a proxy to the non-rotating, Schwarzschild black hole space-time. 
We have ignored the effects of the spin since it affects physical properties 
of the flow very close to the horizon. Thus results from possible 
softest states would be affected by the spin which are not considered here. 
In our case of MAXI~J1543-564 observations, Compton cloud boundary is found to be far away from the 
BH and thus the results are not sensitive to the spin.

We finally conclude that the nature of the evolution of the spectral and temporal properties 
of the source follows a similar trend as observed in other transient BH sources 
(DCM14, MDC14, DMC15, DMCM15, JDCMM16). Detailed study of the complete spectral and 
timing properties of the source is in progress and will be presented else where. 

\section*{Acknowledgements}
D.C. and D.D. acknowledge support from DST sponsored Fast-track Young Scientist project fund (SR/FTP/PS-188/2012).
A.J. and D.D. acknowledge support from ISRO sponsored RESPOND project fund (ISRO/RES/2/388/2014-15).
S.M. acknowledge supports from MoES sponsored post-doctoral research fellowship and FONDECYT post-doctoral grant (\# 3160350).

{}

\clearpage

\begin{figure}
\vskip 0.8cm
        \centerline{
        \includegraphics[scale=0.6,width=8truecm,angle=0]{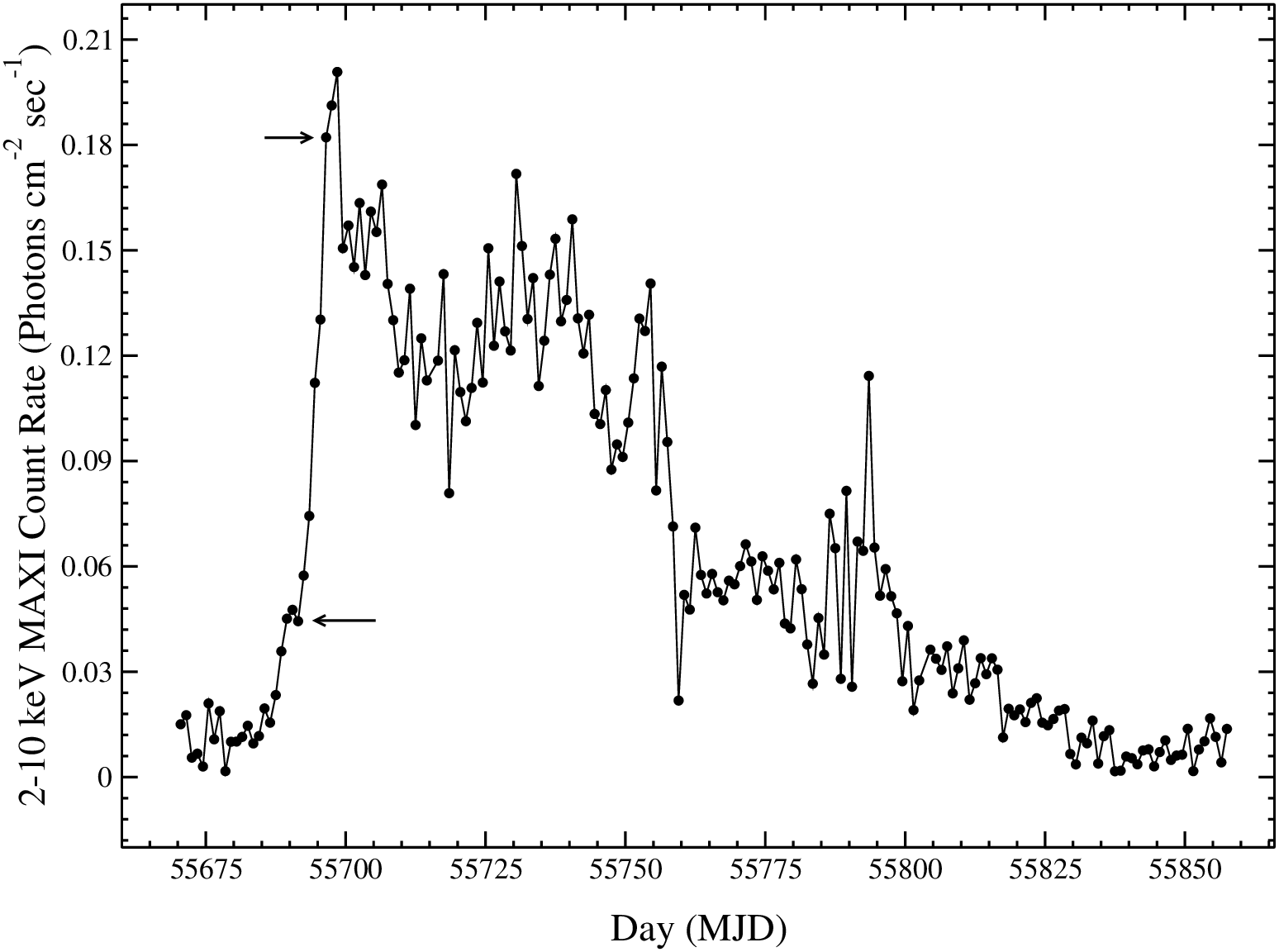}
        }
\caption{Variation of $2-10$~keV MAXI photon count rate in units of photons~cm$^{-2}$~$sec^{-1}$ for the entire 
2011 outburst phase is shown. Period of our RXTE/PCA observations (MJD=55691.09 to 55696.66) presented 
in this paper are marked by the region in between two arrows.}
       \label{fig1}
\end{figure}

\begin{figure}
\vskip 0.8cm
        \centerline{
        \includegraphics[scale=0.6,width=5.8truecm,angle=270]{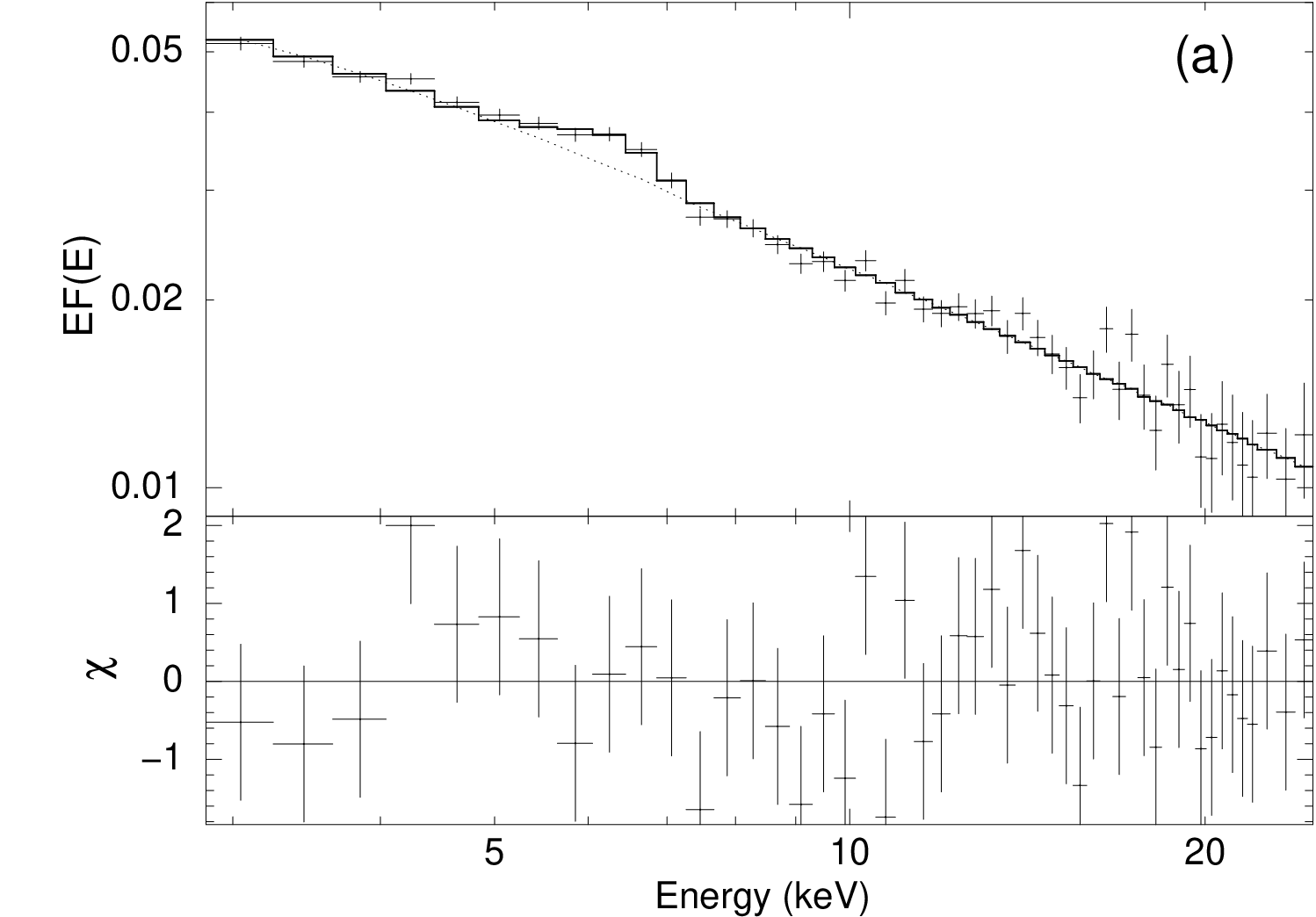}}\vskip 0.3cm
        \centerline{
        \includegraphics[scale=0.6,width=5.8truecm,angle=270]{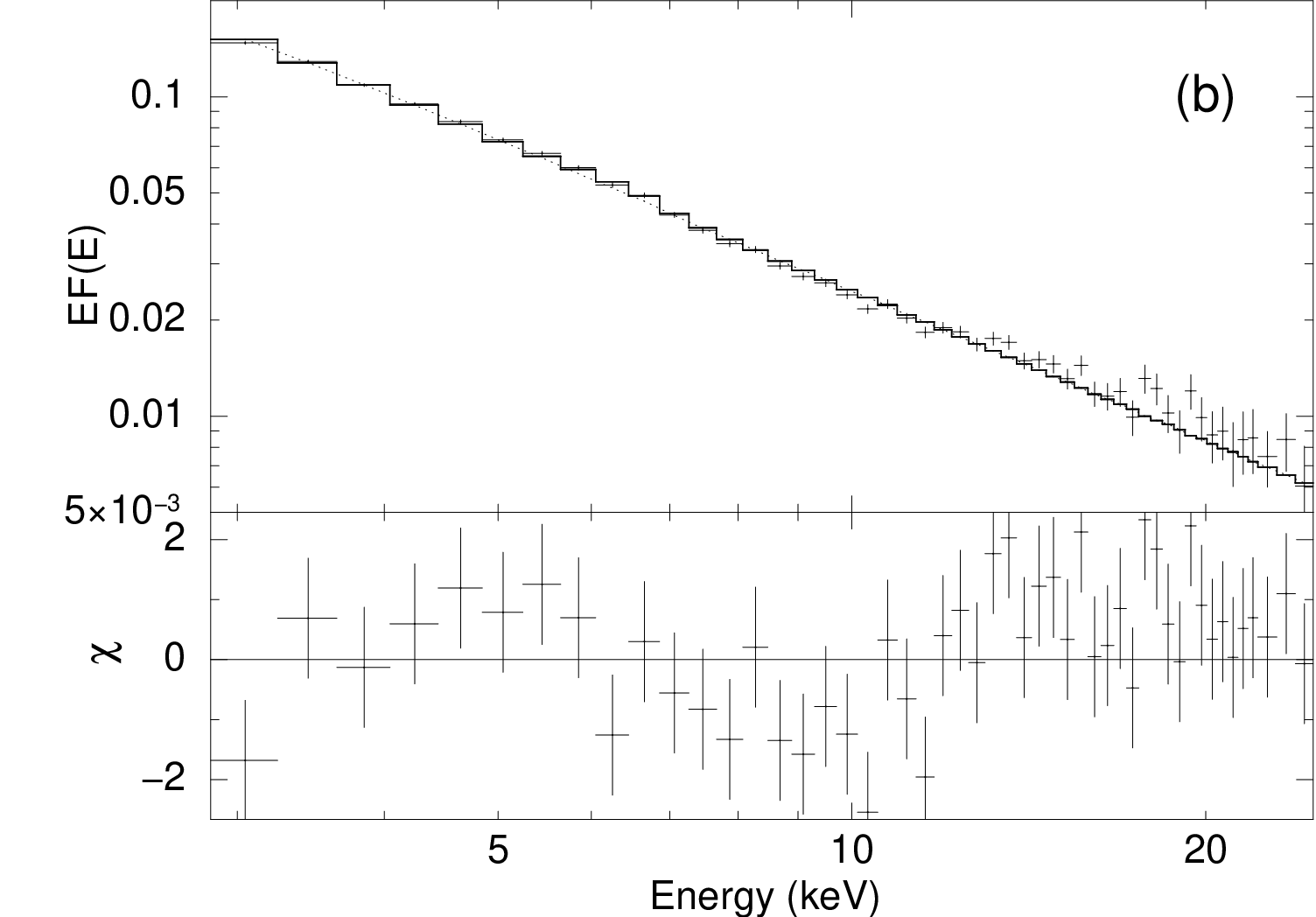}
        }
\caption{TCAF model fitted spectra for the rising (a) HIMS, and (b) SIMS for observations Ids : 
96371-02-01-01 (MJD=55692.09), 96371-02-02-01 (MJD=55694.89) respectively with $\Delta\chi$ 
variations are shown. Note y-axes of the top panels are plotted in E~F(E) with units of keV~(photons~cm$^{-2}$~s$^{-1}$~keV$^{-1}$). }
       \label{fig2}
\end{figure}

\begin{figure}
\vskip 0.8cm
        \centerline{
        \includegraphics[scale=0.6,width=9truecm,angle=0]{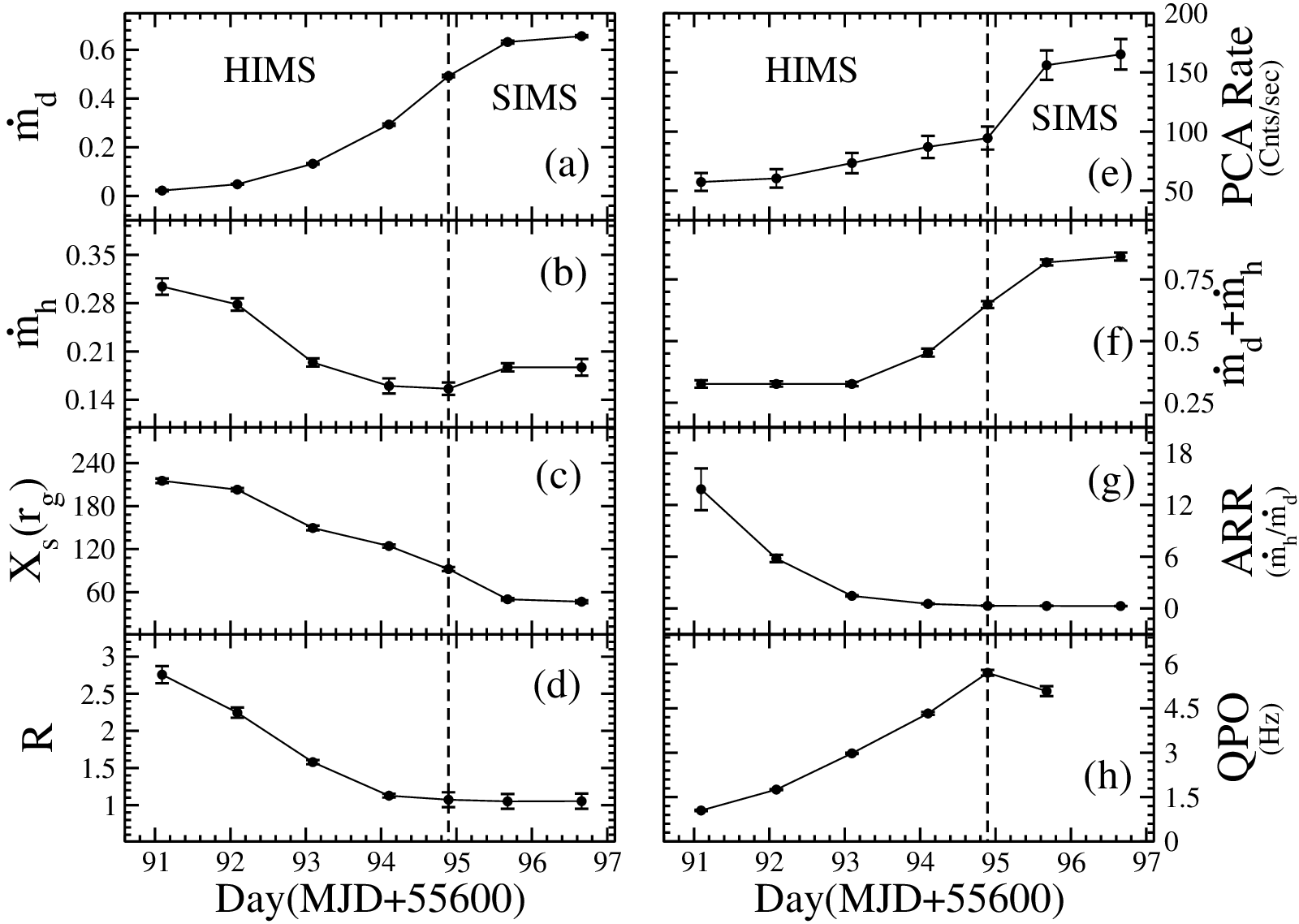}
        }
\caption{Variations of (a) disk rate ($\dot{m_d}$) in $\dot{M}_{Edd}$, (b) halo rate ($\dot{m_h}$) in $\dot{M}_{Edd}$, 
(c) shock location ($X_s$) in $(r_g)$, (d) compression ratio (R), (e) PCU2 count rate in counts per sec, 
(f) total flow rate ($\dot{m_d}+\dot{m_h}$) in $\dot{M}_{Edd}$, (g) ARR, i.e., $\dot{m_h}/\dot{m_d}$, (h) observed 
QPO frequency in Hz with day (in MJD). Note, X-axis markers are modified by subtracting 55600 
from the actual MJDs.}
       \label{fig3}
\end{figure}

\begin{figure}
\vskip 0.8cm
        \centerline{
	\includegraphics[scale=0.6,width=8truecm,angle=0]{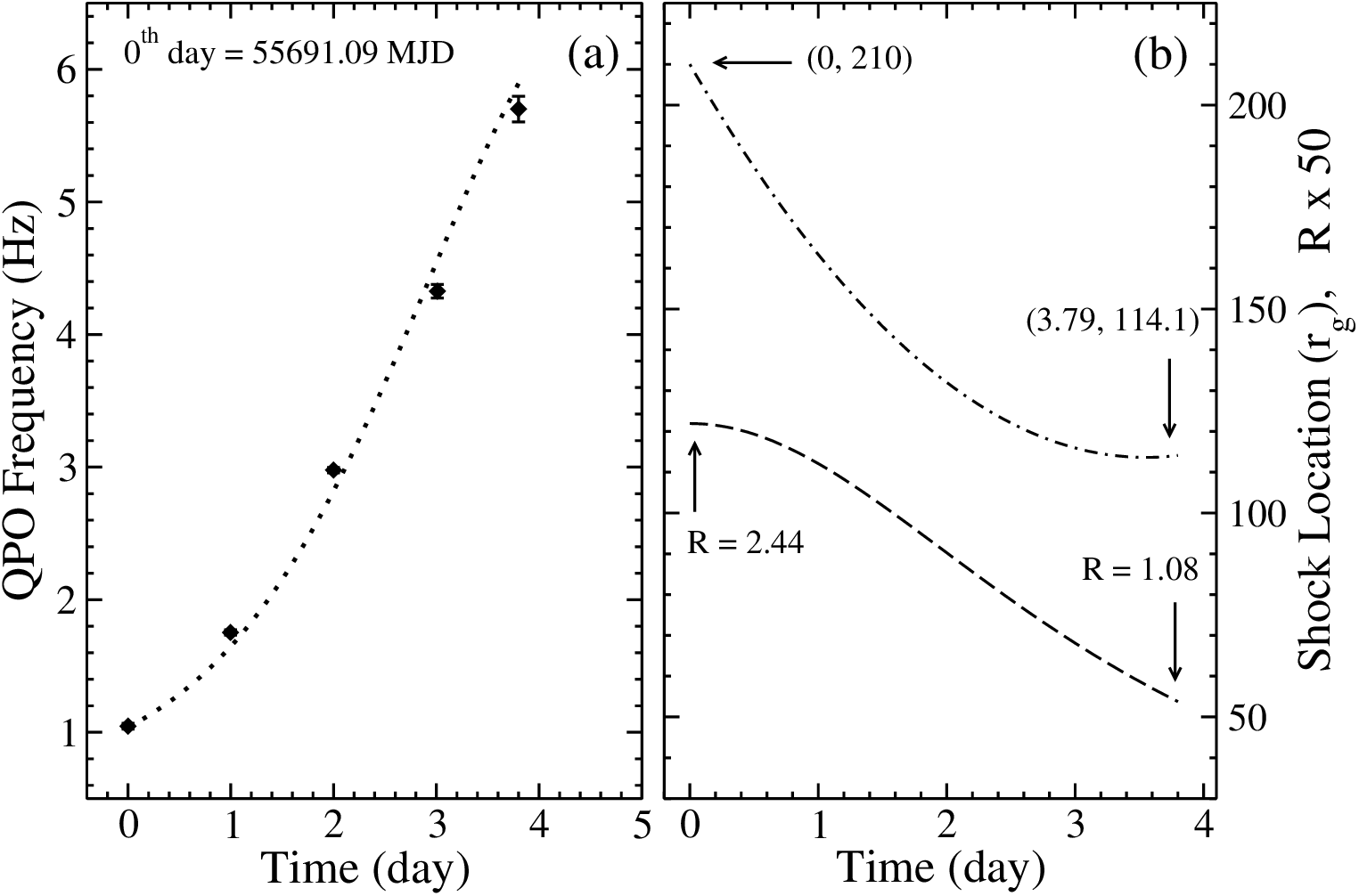}
        }
\caption{(a) Variation of type-C QPO frequency with time (in day) during the rising phase of the 2011 outburst 
of MAXI~J1543-564, fitted with the POS model solution (dashed curve). In right panel (b), variation of 
POS model fitted shock locations (in $r_g$) and compression ratios are shown. }
\label{fig4}
\end{figure}

\begin{figure}
\vskip -0.5cm
        \centerline{
        \includegraphics[scale=0.6,width=6.5truecm,angle=0]{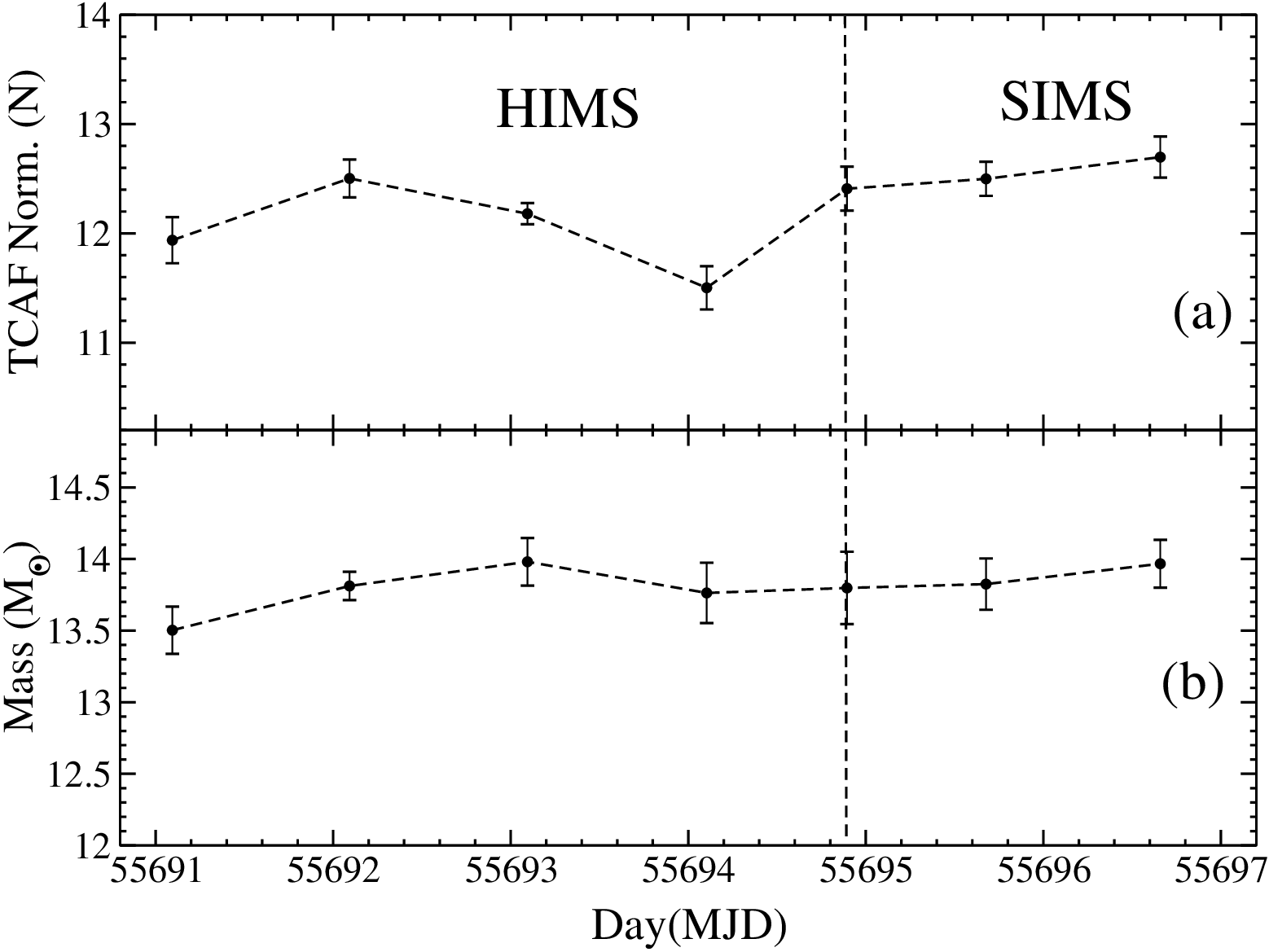}}\vskip 0.3cm
        \centerline{
        \includegraphics[scale=0.6,width=6.5truecm,angle=0]{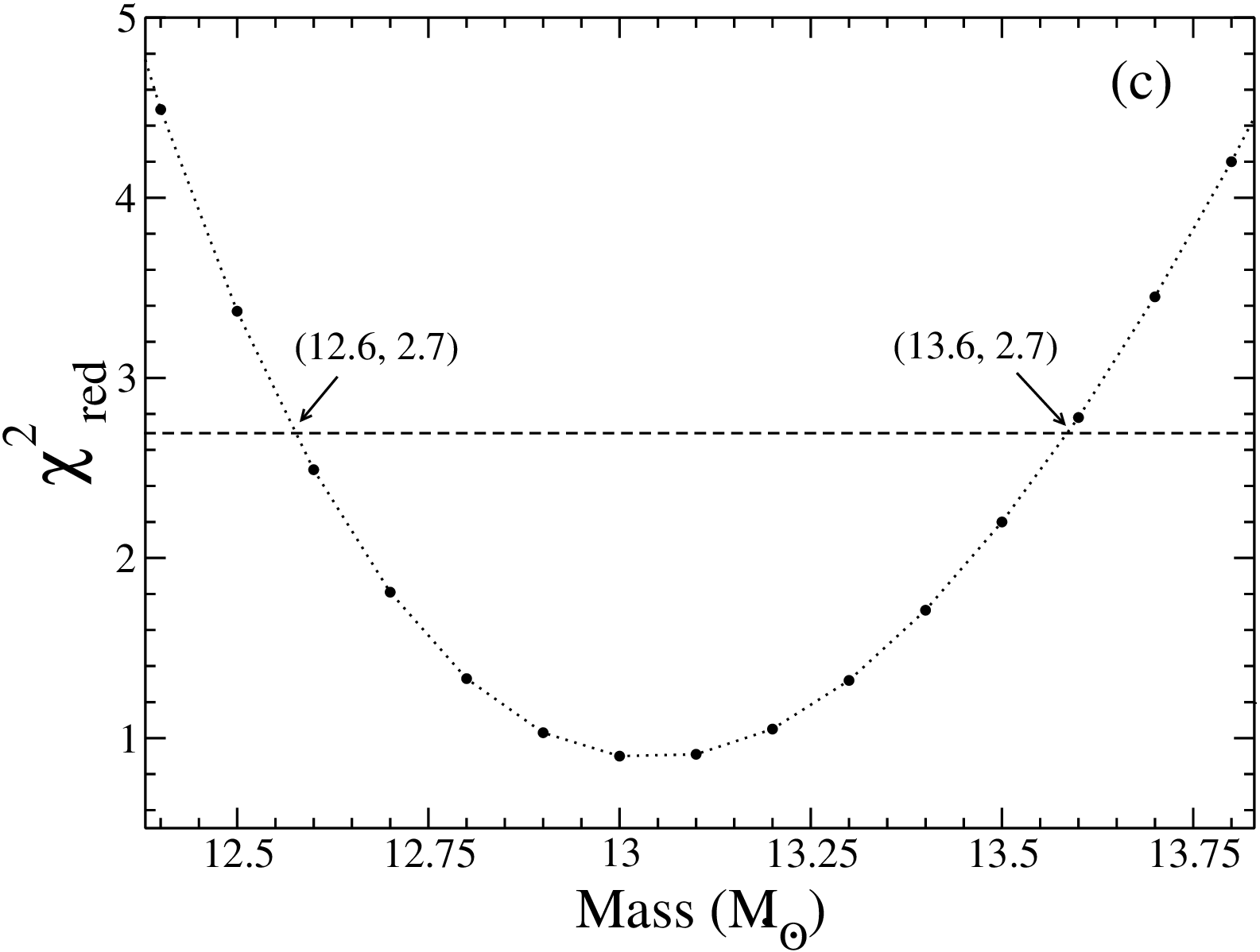}
        }
\caption{Variations of TCAF model fitted (a) normalization (which shows a narrow range of $\sim 11.5-12.8$),
(b) mass of the BH with MJD are shown. we observe a narrow mass range of $\sim 13.5-14.0~M_\odot$. 
In right panel (c) POS model fitted variation of source mass with $\chi^2_{red}$ is shown, which allows 
us to predict mass range if we restrict ourself $\chi^2_{red} \leq 2.7$ for the best fit.}
\label{fig5}
\end{figure}

\clearpage

\begin{table}
\addtolength{\tabcolsep}{-5.0pt}
\centering
\caption{TCAF model fitted spectral parameters}
\label{tab:table1}
\begin{tabular}{lccccccccccc} 
\hline
 $\dot{m_d}$ & $\dot{m_h}$ & ARR  &  $X_s$ & R &   $M_{BH}$      & Norm  & $\chi^2/dof$ \\
  ($\dot{M}_{Edd}$) & ($\dot{M}_{Edd}$) &      &      &  $(r_g)$&   ($M_\odot$)  &      &   \\
(1)   & (2) & (3) & (4) & (5) & (6) & (7) & (8)\\ 
\hline
 $0.022^{\pm 0.003}$& $0.304^{\pm 0.012}$&$13.8^{\pm 2.4 }$& $215.4^{\pm 3.2}$& $2.76^{\pm 0.12}$& $13.5^{\pm 0.2}$&$12.0^{\pm 0.2}$&41.62/41 \\
 $0.048^{\pm 0.002}$& $0.278^{\pm 0.009}$&$5.79^{\pm 0.43}$& $203.0^{\pm 2.1}$& $2.25^{\pm 0.07}$& $13.8^{\pm 0.1}$&$12.5^{\pm 0.2}$&42.61/41 \\
 $0.132^{\pm 0.003}$& $0.194^{\pm 0.006}$&$1.47^{\pm 0.08}$& $149.6^{\pm 3.3}$& $1.58^{\pm 0.03}$& $13.9^{\pm 0.2}$&$12.2^{\pm 0.1}$&66.86/41 \\
 $0.293^{\pm 0.009}$& $0.160^{\pm 0.011}$&$0.55^{\pm 0.05}$& $124.4^{\pm 2.2}$& $1.13^{\pm 0.03}$& $13.8^{\pm 0.2}$&$11.5^{\pm 0.2}$&60.92/41 \\
 $0.492^{\pm 0.013}$& $0.156^{\pm 0.009}$&$0.32^{\pm 0.03}$& $ 92.2^{\pm 2.7}$& $1.07^{\pm 0.10}$& $13.8^{\pm 0.3}$&$12.4^{\pm 0.2}$&67.20/41 \\
 $0.632^{\pm 0.021}$& $0.187^{\pm 0.006}$&$0.29^{\pm 0.02}$& $ 50.0^{\pm 1.7}$& $1.05^{\pm 0.10}$& $13.8^{\pm 0.2}$&$12.5^{\pm 0.2}$&69.34/41 \\
 $0.656^{\pm 0.019}$& $0.186^{\pm 0.012}$&$0.29^{\pm 0.03}$& $ 46.6^{\pm 1.9}$& $1.05^{\pm 0.10}$& $14.0^{\pm 0.3}$&$12.7^{\pm 0.2}$&69.92/41 \\
\hline
\end{tabular}
\noindent{
\leftline{$\dot{m_d}$ and $\dot{m_h}$ are the TCAF model fitted disk rate and halo rate (both in $\dot{M}_{Edd}$).}
\leftline{ARR (accretion rate ratio) is the ratio between $\dot{m_h}$ and $\dot{m_d}$, i.e., $\dot{m_h}/\dot{m_d}.$}
\leftline{Shock location ($X_s$) is in $r_g$ and mass of the BH ($M_{BH}$) is in $M_{\odot}$.}
\leftline{$dof$ is degrees of freedom, the ratio of $\chi^2$ and $dof$ is the $\chi^2_{red}$.}
\leftline{Note: average values of 90\% confidence `$\pm$' parameter errors are mentioned in superscripts.}
}
\end{table}

\begin{table}
\addtolength{\tabcolsep}{-3.50pt}
\centering
\caption{QPO evolution in initial rising phase: Fitted with POS Model}
\label{tab:table2}
\begin{tabular}{lccccccccc} 
\hline
 Obs. & Id.&  MJD & $\nu_{Obs}$ & $\nu_{POS}$ & $X_s$   & V       & R & $\nu_{TCAF}$\\
      &    &          & (Hz)    &   (Hz)      & $(r_g)$ &(cm/s)   &   & (Hz) \\
(1)   & (2) & (3) & (4) & (5) & (6) & (7) & (8) & (9) \\  
\hline
1 & X-01-00& 55691.09 & 1.05$^{\pm 0.02}$ & 1.04 & 210.0 & 2450.0 & 2.44& 0.85$^{\pm 0.06}$\\
2 & X-01-01& 55692.09 & 1.75$^{\pm 0.02}$ & 1.65 & 163.4 & 2105.7 & 2.24& 1.12$^{\pm 0.05}$\\
3 & X-01-02& 55693.09 & 2.98$^{\pm 0.02}$ & 2.82 & 132.0 & 1760.1 & 1.81& 2.49$^{\pm 0.14}$\\
4 & X-02-00& 55694.10 & 4.38$^{\pm 0.05}$ & 4.56 & 115.9 & 1411.5 & 1.36& 4.67$^{\pm 0.27}$\\
5 & X-02-01& 55694.89 & 5.70$^{\pm 0.09}$ & 5.89 & 114.1 & 1139.0 & 1.08& 7.68$^{\pm 1.11}$\\
6*& X-02-02& 55695.67 & 5.08$^{\pm 0.17}$ & ---- & ---- & ----  & ----& ----\\ 
7 & X-02-03& 55696.68 & ----- & ---- & ---- & ----  & ----& ----\\ 
\hline
\end{tabular}
\noindent{
\leftline{Here 'X'=96371-02 signifies the initial part of an observation Id.}
\leftline{Here, $\nu_{Obs}$ is the observed QPO frequency, $\nu_{POS}$ is the theoretical QPO}
\leftline{frequency calculated from the POS model fit, `V' is the velocity of the shock in $cm~sec^{-1}$,}
\leftline{`$R$' is the shock compression ratio, and $\nu_{TCAF}$ is the calculated QPO}
\leftline{frequency from TCAF model fitted shock parameters.} 
\leftline{$^*$ This type-B QPO of SIMS does not fitted with the POS, since origin of the type-B QPOs are different.}
}
\end{table}


\begin{thebibliography}{99}

\bibitem[Altamirano et al.(2011)]{Altamirano11}Altamirano, D., Kalamkar, M., Yang, Y., et al., 2011, ATel 3334,1
\bibitem[Arnaud(1996)]{Arnaud96} Arnaud, K.A., 1996, ASP Conf. Ser., Astronomical Data Analysis Software and Systems V, ed. G.H. Jacoby \& J. Barnes, 101, 17
\bibitem[Casella et al.(2005)]{CA05} Casella, P., Belloni, T., Stella, L., 2005, ApJ 629,403
\bibitem[Chakrabarti \& Titarchuk(1995)]{CT95} Chakrabarti, S.K., \& Titarchuk, L.G., 1995, ApJ, 455, 623 (CT95)
\bibitem[Chakrabarti \& Manickam (2000)]{C00}Chakrabarti, S.K. \& Manickam, S.G., 2000, ApJ, 531, L41
\bibitem[Chakrabarti(1997)]{C97} Chakrabarti, S.K., 1997, ApJ, 484, 313
\bibitem[Chakrabarti et al.(2005)]{C05}Chakrabarti, S.K., Nandi, A., \& Debnath, D., et al., 2005, IJP, 79, 841 (arXiv:astro-ph/0508024)
\bibitem[Chakrabarti et al.(2008)]{C08} Chakrabarti, S.K., Debnath, D., \& Nandi, A. et al., 2008, A\&A, 489, L41
\bibitem[Chakrabarti et al. (2015)]{C15} Chakrabarti S.K., Mondal, S., Debnath, D., 2015, MNRAS, 452, 3451
\bibitem[Debnath et al.(2008)]{DD08} Debnath, D., Chakrabarti, S.K., Nandi, A., \& Mandal, S., 2008, BASI, 36, 151
\bibitem[Debnath et al.(2010)]{DD10}Debnath, D., Chakrabarti, S. K., \& Nandi, A., 2010, A\&A, 520, 98
\bibitem[Debnath et al.(2013)]{DD13} Debnath, D., Chakrabarti, S.K., \& Nandi, A., 2013, AdSpR, 52, 2143
\bibitem[Debnath, Chakrabarti \& Mondal(2014)]{DCM14} Debnath, D., Chakrabarti, S.K., \& Mondal, S., 2014, MNRAS, 440, L121 (DCM14)
\bibitem[Debnath, Mondal \& Chakrabarti(2015a)]{DMC15} Debnath, D., Mondal, S., \& Chakrabarti, S.K., 2015a, MNRAS, 447, 1984 (DMC15)
\bibitem[Debnath et al.(2015b)]{DMCM15} Debnath, D., Molla, A. A., Chakrabarti, S.K., \& Mondal, S., 2015b, ApJ, 803, 59 (DMCM15)
\bibitem[Jana et al.(2015)]{Jana15} Jana, A., Debnath, D., \& Chakrabarti, S. K., et al., 2016, ApJ, 819, 107 (JDCMM16)
\bibitem[Kennea et al.(2011)]{Kennea11a}Kennea, J. A., Evans, P. A., Krimm, H. A., et al., 2011, ATel 3331,1
\bibitem[Kennea et al.(2011)]{Kennea11b}Kennea, J., Esposito, P., Israel, G., et al., 2011, ATel 3336,1
\bibitem[McClintock \& Remillard(2006)]{MR06}McClintock, J. E., \& Remillard, R. A., 2006, in Compact Stellar X-ray Sources, Ed. W. Lewin \& M. van der Klis, 39, 157
\bibitem[Miller-Jones et al.(2011)]{Miller11}Miller-Jones, J.C.A., Tzioumis, A.K., Jonker, P.G., et al., 2011, ATel 3364,1
\bibitem[Molla et al.(2016)]{Molla16} Molla, A. A., Debnath, D., \& Chakrabarti, S. K., et al., 2016, MNRAS (in press) (arXiv:1605.07282)
\bibitem[Molteni et al.(1996)]{MSC96}Moltani, D., Sponholz, H., \& Chakrabarti, S.K., 1996, ApJ, 457, 805
\bibitem[Mondal, Chakrabarti \& Debnath(2015)]{Mondal15} Mondal, S., Chakrabarti, S.K., \& Debnath, D., 2015, ApJ, 798, 57
\bibitem[Mondal, Debnath \& Chakrabarti(2014)]{Mondal14} Mondal, S., Debnath, D., \& Chakrabarti, S.K., 2014, ApJ, 786, 4 (MDC14)
\bibitem[Munoz-Darias et al.(2011)]{Munoz11}Munoz-Darius, T., Motta, S., Stiele, H., et al., 2011, ATel 3341,1
\bibitem[Nandi et al.(2012)]{Nandi12} Nandi, A., Debnath, D., Mandal, S., \& Chakrabarti, S.K., 2012, A\&A, 542, A56
\bibitem[Negoro et al.(2011)]{Negoro11}Negoro, H., Nakahira, S., Ueda, Y., et al., 2011, ATel 3330,1
\bibitem[Paczy\'nski \& Wiita (1980)]{PW80}Paczy\'nski, B. \& Wiita, P.J., 1980, A\&A, 88, 23
\bibitem[Rapisarda et al. (2013)]{Rap13}Rapisarda, S., Ingram, A. \& van der Klis, M., 2013, MNRAS 
\bibitem[Rau et al. (2011)]{Rau11}Rau, A., Greiner, J., Elliot, J., et al., 2011, Atel 3365,1
\bibitem[Remillard \& McClintock(2006)]{RM06} Remillard, R.A., \& McClintock, J.E., 2006, ARA\&A, 44, 49
\bibitem[Rojas et al. (2011)]{Rojas11}Rojas, A.F., Masetti, N., Minniti, D., 2011, Atel 3372,1
\bibitem[Russell et al. (2011)]{Russel11}Russell, D., Lewis, F., Roche, P., et al., 2011, ATel 3359,1
\bibitem[Ryu et al. (1997)]{RCM97}Ryu, D., Chakrabarti, S. K., \& Molteni, D. 1997, ApJ, 474, 378
\bibitem[Shakura \& Sunyaev(1973)]{SS73} Shakura, N.I., \& Sunyaev, R.A., 1973, A\&A, 24, 337 (SS73)
\bibitem[Stiele et al.(2011)]{Stiele11}Stiele, H., Munoz-Darias, T., Motta, S., \& Belloni, T.M., 2011, MNRAS
\bibitem[Sunyaev \& Titarchuk (1980)]{ST80}Sunyaev, R.A., \& Titarchuk, L.G., 1980, ApJ, 86, 121
\bibitem[Sunyaev \& Titarchuk (1985)]{ST85}Sunyaev, R.A., \& Titarchuk, L.G., 1985, A\&A, 143, 374
\bibitem[Tomsick et al.(2014)]{Tomsick14}Tomsick, J. A., Yamaoka, K., \& Corbel, S., et al., 2014, ApJ, 791, 70

\end{thebibliography}
\end{document}